\documentclass[conference]{IEEEtran}
\IEEEoverridecommandlockouts
\usepackage{cite}
\usepackage{amsmath,amssymb,amsfonts}
\usepackage{algorithmic}
\usepackage{graphicx}
\usepackage{textcomp}
\usepackage{textgreek}
\usepackage{float}
\usepackage{xcolor}
\def\BibTeX{{\rm B\kern-.05em{\sc i\kern-.025em b}\kern-.08em
    T\kern-.1667em\lower.7ex\hbox{E}\kern-.125emX}}
\begin{document}

\title{Electrical Impedance Tomography with an Integrated Picoliter-Volume Subtractive Microfluidic Chamber in 65 nm CMOS\\
\thanks{This work was supported by the Caltech Summer Undergraduate Research Fellowship Program, National Science Foundation Graduate Research Fellowship under Grant No. DGE-1745301, the Space Solar Power Project, and the Carver Mead New Adventures Fund.}
}

\author{\IEEEauthorblockN{Antônio Victor Machado de Oliveira, Debjit Sarkar, and Ali Hajimiri}
\IEEEauthorblockA{\textit{Department of Electrical Engineering} \\
\textit{California Institute of Technology}\\
Pasadena, California, USA \\
Email: amachado@caltech.edu}
}

\maketitle

\begin{abstract}

Electrical impedance tomography with fully integrated microfluidics and electronics is presented for the first time in a CMOS chip. Chambers and electrodes are fabricated in the interconnect layers of a 65 nm CMOS chip through post-processing, enabling picoliter-volumes to be processed and imaged. Tomography maps are reconstructed by reading out voltages from a 16-element electrode array and processing the data off-chip, and sources of variation in reconstruction are discussed. The EIT system presented in this work serves as a proof-of-concept towards using CMOS as a platform for co-integrated microfluidics and electronics. 

\end{abstract}


\section{Introduction}

\begin{figure}[htbp]
\centering
\includegraphics[width=0.55\linewidth]{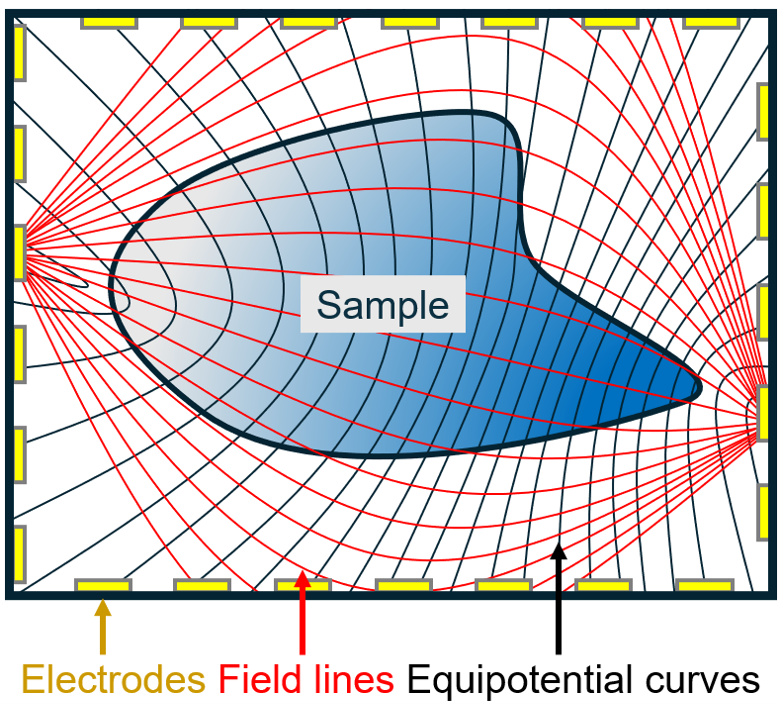}
\caption{Illustration of field lines and equipotential curves for a pair of electrodes.}
\label{eit_intro}
\end{figure}

Electrical impedance tomography (EIT) is a non-invasive, label-free imaging technique for biological systems \cite{neweit, hybrideit, miniatureeit}. The technique requires the injection of currents and measurement of boundary voltages to reconstruct conductivity maps of the sample, as illustrated in Fig. \ref{eit_intro}. It has attracted interest for applications like brain and cardiac imaging, probing cell cultures, and even monitoring single-cell behavior due to close links between dielectric properties and physiological state \cite{neweit, activeelec}. These applications motivate miniaturizing EIT systems to be compatible with microscale biological environments, with EIT platforms in CMOS potentially offering high electrode densities, simplified packaging and sample confinement, and on-chip processing using integrated electronics \cite{tactileeit, drugs, scaffold}.



Despite these advantages, miniaturized EIT remains technically challenging, as smaller volumes increase sensitivity to imperfections and impurities \cite{hybrideit}. Additionally, the inverse problem of EIT is fundamentally ill-posed, with different forms of regularization necessary to stabilize solutions in state-of-the-art reconstruction algorithms \cite{bayesian}. To address these challenges, we present a fully integrated EIT system in 65 nm CMOS featuring a picoliter-volume  microfluidic chamber, digitally reconfigurable electrodes, and circuitry for stimulation and sensing.

%





\section{Integrated Microfluidics}

The microfluidics described in this work are fully integrated inside the foundry-fabricated CMOS chip. Components are created by selectively etching metal in the interconnect layers of the chip, leaving behind structures made of inter-metal dielectrics (IMDs) \cite{subtractive, oldsubtractive}. These IMDs largely consist of silicon oxides and nitrides, which are suitable materials for interfacing with biological samples \cite{visiblespectrum}. In prior literature, this subtractive method has primarily been used to integrate photonic circuits within electronic CMOS chips without needing to modify the semiconductor manufacturing process \cite{subtractive}. In this work, we show that this method also enables the fabrication of microfluidic components without the addition of materials like silicon or plastic.

\begin{figure}[htbp]
\centering
\includegraphics[width=0.9\linewidth]{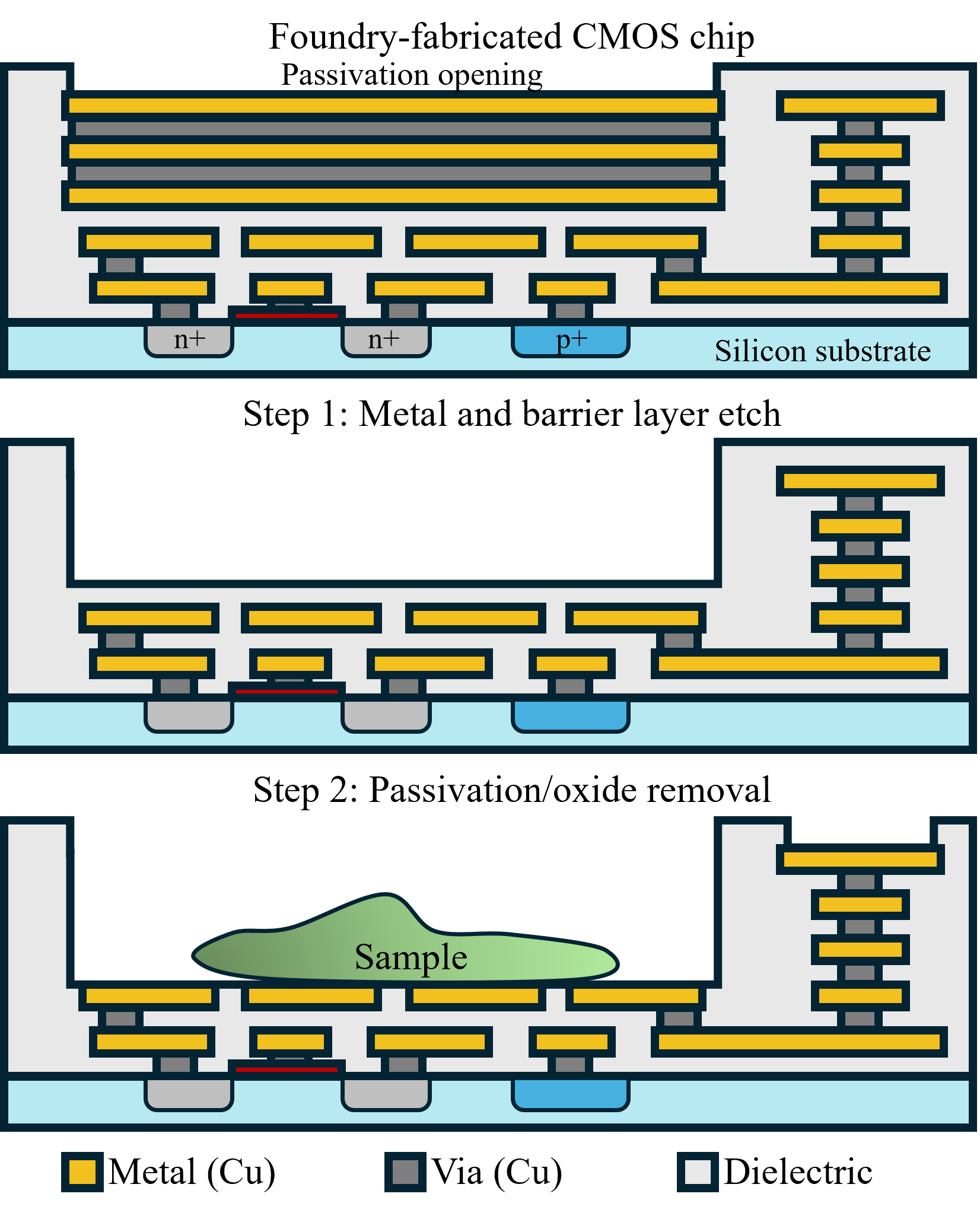}
\caption{Simplified illustration of the subtractive method used to create microfluidic devices in CMOS. The metal and barrier layers are removed by applying Al Etch Type A and EDTA/H$_2$O$_2$ in an alternating manner. Passivation and oxides are removed via laser ablation.}
\label{etching_process}
\end{figure}

An illustration of the fabrication procedure can be seen in Fig. \ref{etching_process}. A standard CMOS chip is designed with passivation covering all metals except those that will be etched. The exposed metals are then wet-etched, with Aluminum Etch Type A being used to remove metals like copper and aluminum and a EDTA/H$_2$O$_2$ mixture used to remove barrier layers like Ta/TaN or Ti/TiN \cite{oldsubtractive}. Finally, the passivation is exposed through laser ablation so that electronic connections can be made to the chip. If electrodes need to be exposed, such as in the case of the EIT system described in this work, laser ablation exposes them as well.

\begin{figure}[htbp]
\includegraphics[width=\linewidth]{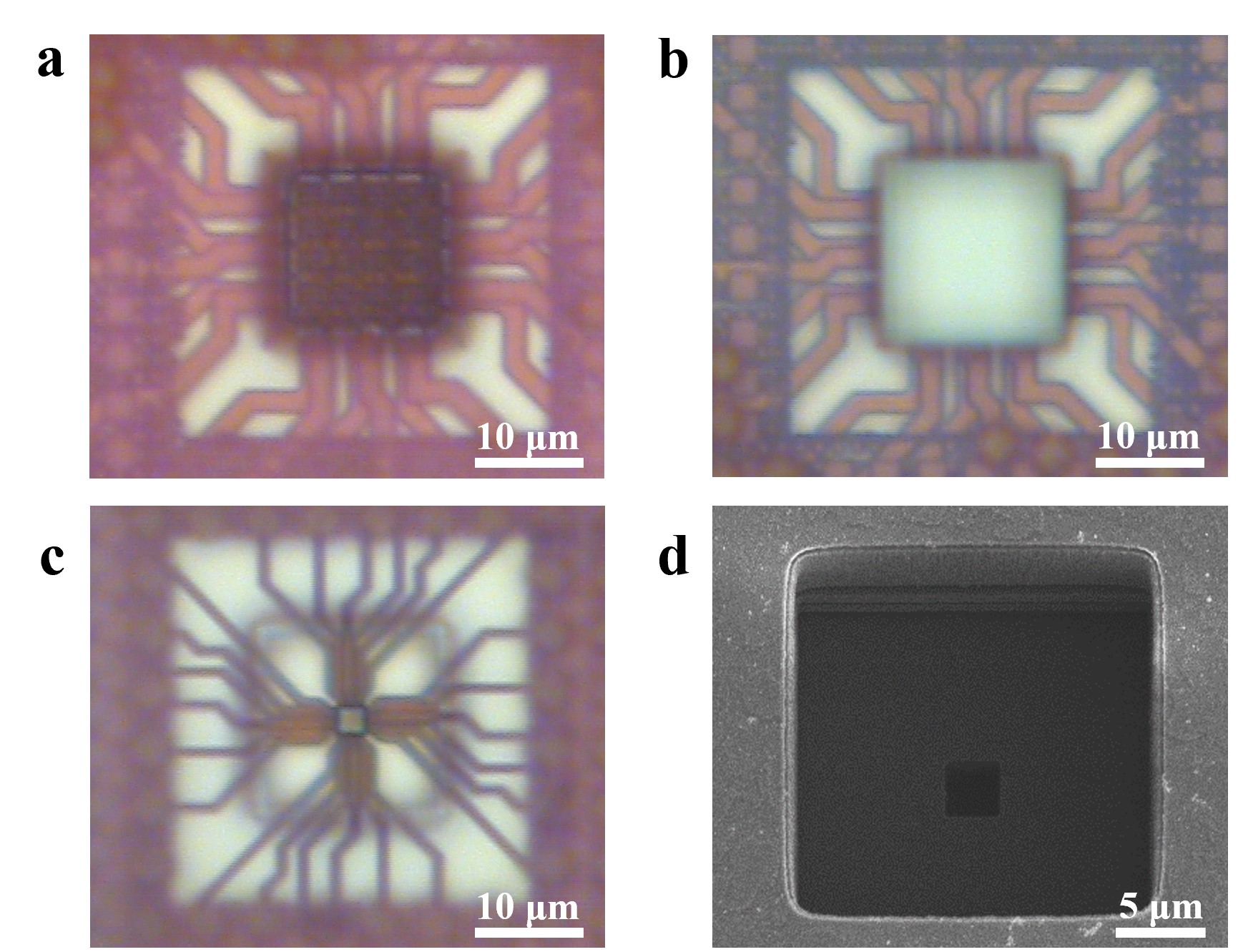}
\caption{Micrograph of a 20 \textmu m by 20 \textmu m microfluidic chamber (a) before and (b) after etching. (c) Micrograph of a 20 \textmu m by 20 \textmu m chamber with 3 \textmu m by 3 \textmu m chamber inside and (d) corresponding SEM image.}
\label{eit_chambers_old}
\end{figure}

Examples of test-structures fabricated using this method can be seen in Fig. \ref{eit_chambers_old}. The chambers pictured in Fig. \ref{eit_chambers_old}a and \ref{eit_chambers_old}b have a capacity of 4.9 pL, while the chambers in Fig. \ref{eit_chambers_old}c and \ref{eit_chambers_old}d have a capacity of 3.3 pL for the upper chamber and a capacity of 20 fL for the inset 3 \textmu m by 3 \textmu m section. Electrodes face the chamber on multiple metal layers from all four sides.

\section{System Design}

The fabricated chip features an integrated EIT chamber and circuits for stimulation, sensing, local oscillator (LO) generation, and digital control. The chamber has lateral dimensions of 150 \textmu m by 150 \textmu m and extends approximately 10 \textmu m below the surface of the chip, with a few metal layers below the chamber left unetched for electrodes and electrical connections. The capacity of the chamber is approximately 225 pL, and a grid of 16 electrodes spans the bottom of the chamber with a fill factor of 66\%.

\begin{figure}[htbp]
\includegraphics[width=\linewidth]{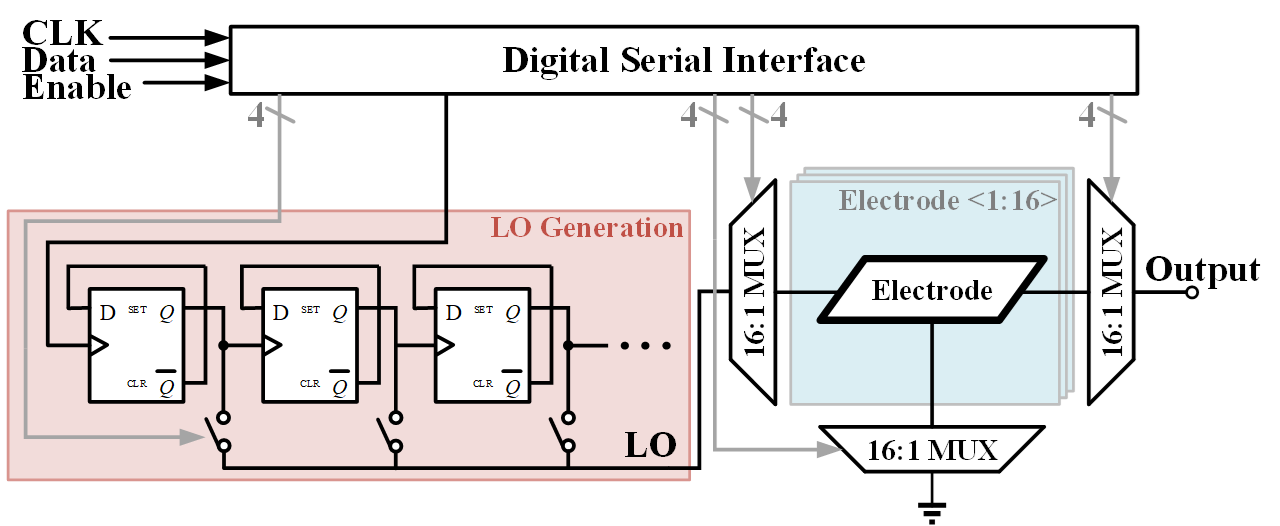}
\caption{Block diagram of the CMOS chip. Electrodes are digitally reconfigured to be connected to the LO, ground, output, or be floating. The output is processed off-chip using EIDORS.}
\label{system_block}
\end{figure}

A diagram of the system can be seen in Fig. \ref{system_block}. The clock signal used to operate the digital serial interface is reused to generate the LO through a series of tunable frequency dividers. The LO is applied across a pair of electrodes, and the voltage is then measured at all other electrodes and sent off-chip for processing. EIDORS, an open-source software for modeling EIT, is used for image reconstruction \cite{eidors}. Multiplexers are used to digitally reconfigure electrode connections for stimulation and sensing.

\section{Measurement Results}
\begin{figure}[htbp]
\centering
\includegraphics[width=0.9\linewidth]{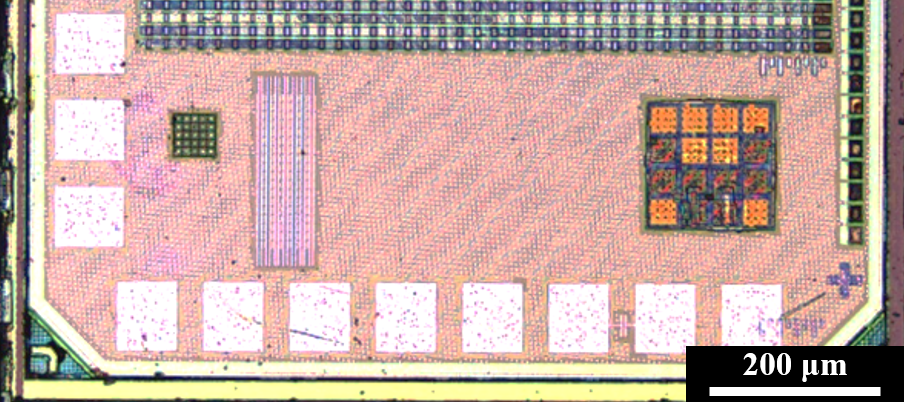}
\caption{Micrograph of the EIT chip.
}
\label{die_photo}
\end{figure}

The EIT chip is fabricated in a 65 nm CMOS process (Fig. \ref{die_photo}). The chip is etched with Aluminum Etch Type A at 80$^\circ$C to remove copper and EDTA and H$_2$O$_2$ at 50$^\circ$C to remove the Ta/TaN barrier bilayer. These etchants are alternated until the microfluidic chamber is fully exposed, and megasonic cleaning is used to clean the chamber in preparation for measurement. The chip operates off of a 1 V supply and consumes 96 to 103 \textmu W for LO frequencies between 500 Hz and 100 kHz, respectively.

\begin{figure}[htbp]
\centering
\includegraphics[width=\linewidth]{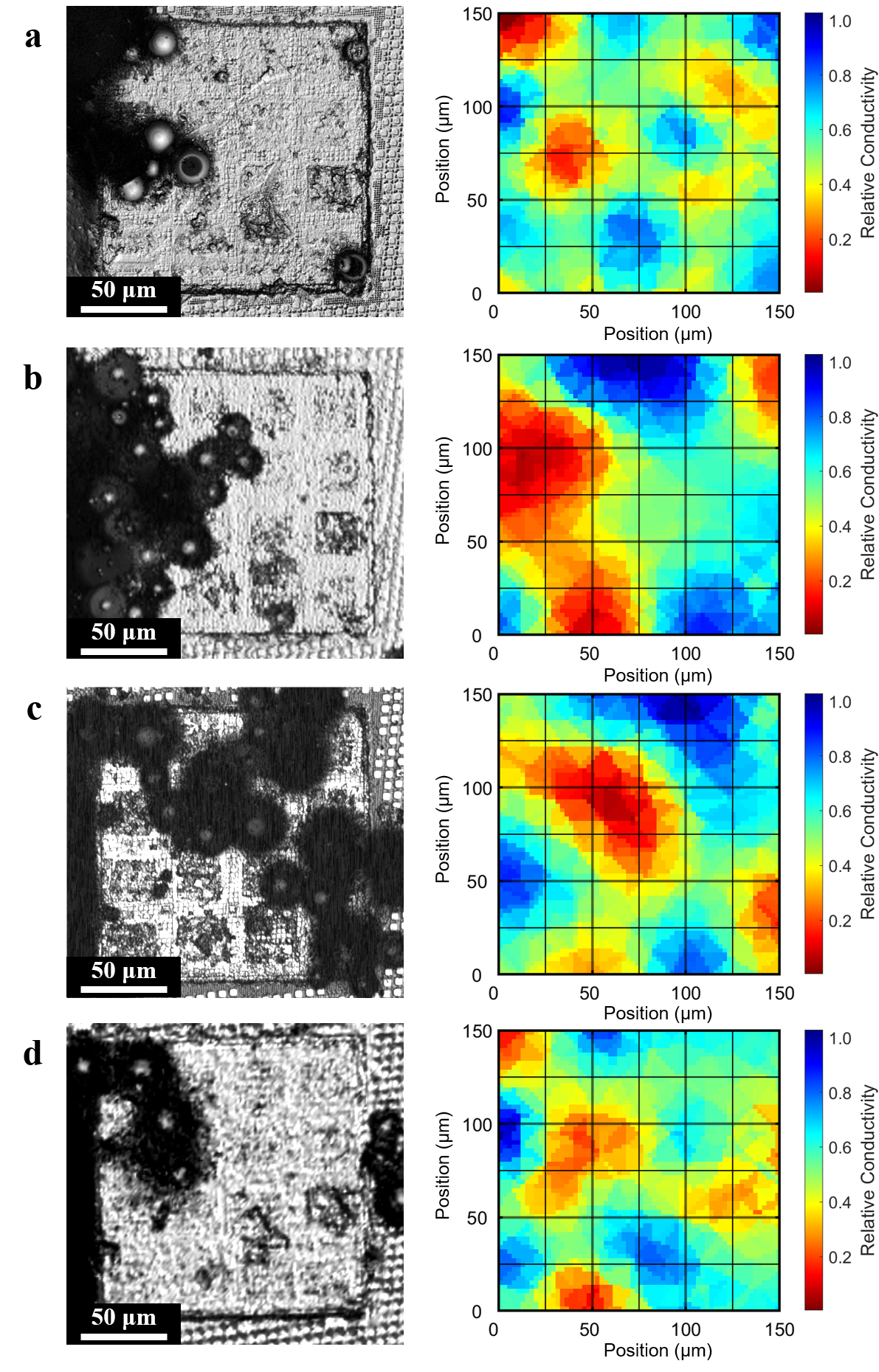}
\caption{Micrographs of samples and their corresponding reconstructed impedance tomography maps. All maps are measured at a frequency of 50 kHz.
}
\label{tomography}
\end{figure}

The performance of the system is evaluated by inserting glass microspheres in a water-filled chamber and relaying the measured signals to EIDORS for tomography reconstruction. Conductivity maps for a variety of configurations can be seen in Fig. \ref{tomography}, and comparing them to images captured under a microscope indicates strong correlation between microsphere positions and areas of low conductivity. Inaccuracies are largely attributed to the ill-posed nature of the EIT problem and the small volume causing increased sensitivity to impurities in the chamber.

\begin{figure}[htbp]
\includegraphics[width=\linewidth]{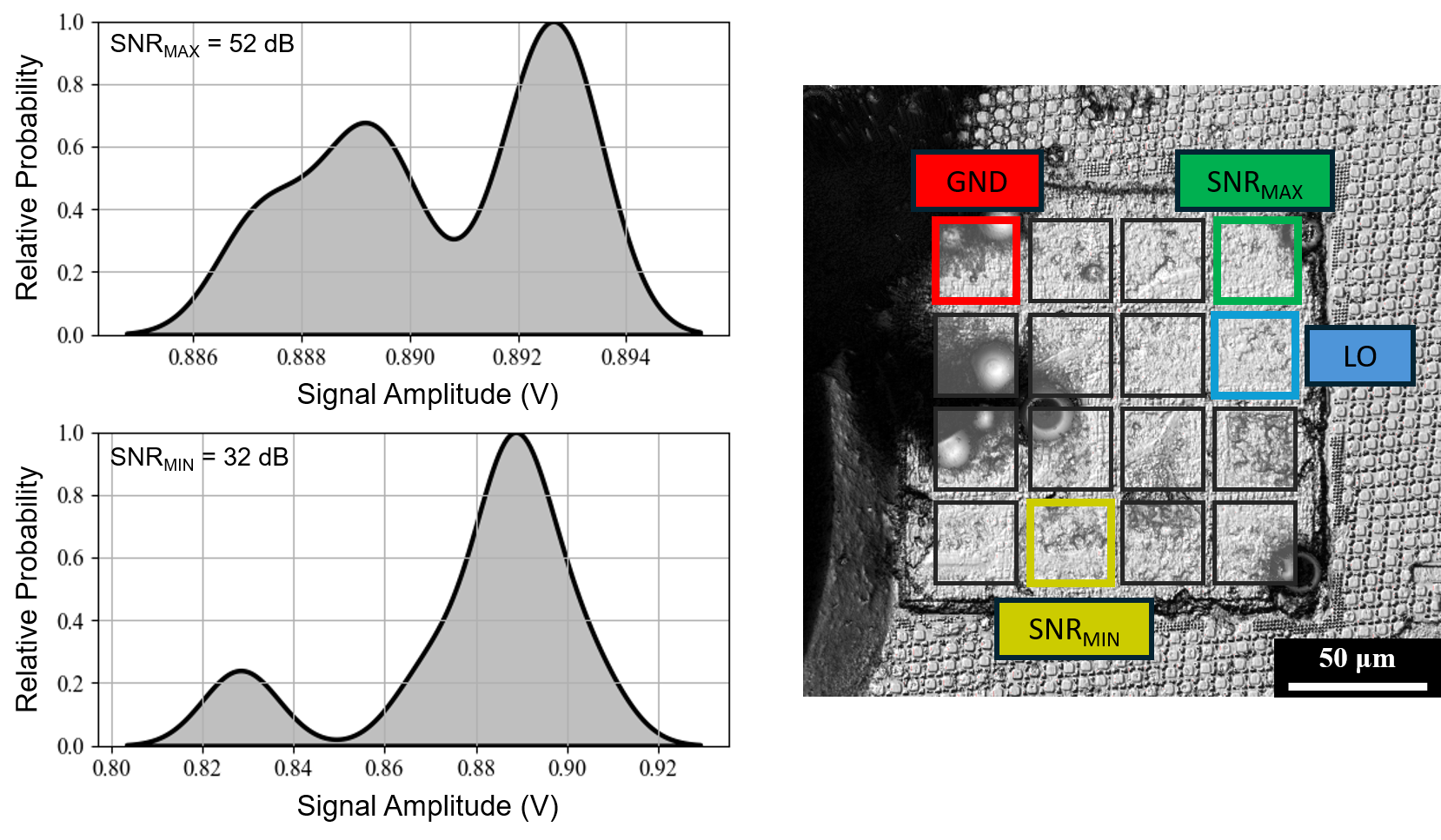}
\caption{Histograms of amplitude variation over time for electrodes measured in Fig. \ref{tomography}a. The electrode positions are annotated on the micrograph, and the electrodes where minimum and maximum SNR are measured are highlighted. Samples are taken continuously over approximately 4 minutes.
}
\label{distribution_plot}
\end{figure}

\begin{figure}[htbp]
\includegraphics[width=\linewidth]{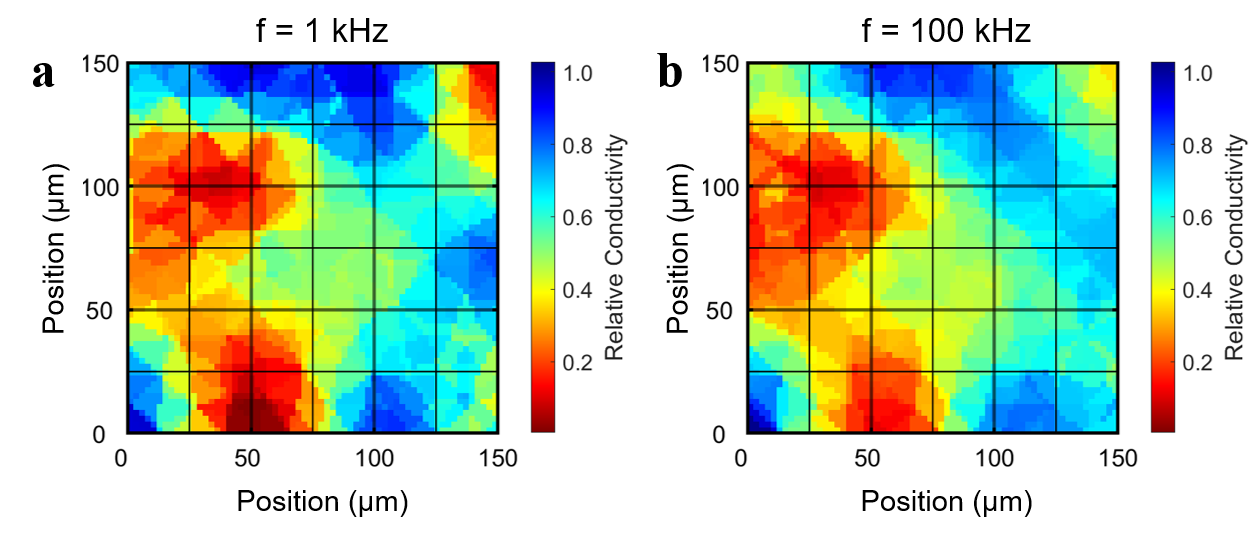}
\caption{Tomography map reconstructed for the sample measured in Fig. \ref{tomography}b at LO frequencies of (a) 1 kHz and (b) 100 kHz.}
\label{vsfreq}
\end{figure}

The signal-to-noise ratio (SNR) of the measurement depends on multiple factors, including the exact sample being measured, the location of the sensing electrode with respect to the stimulated electrodes, and the LO frequency. As an example, the SNR varies between 32 dB and 52 dB for the sample configuration depicted in Fig. \ref{tomography}a, with SNR typically increasing with closer proximity between stimulating and sensing electrodes (Fig. \ref{distribution_plot}). We attribute the shape of the histograms primarily to shifts in local temperature and fluid flow in the chamber. An example of variation versus frequency is shown for the sample configuration in Fig. \ref{tomography}b (Fig. \ref{vsfreq}). These images are fairly similar across frequency, which is expected given that the microspheres maintain a low conductivity versus frequency.



\section{Conclusion}
This work represents, to the best of the authors' knowledge, the first demonstration of electrical impedance tomography with on-chip microfluidics. A picoliter-scale EIT chamber is fabricated in the interconnect layers of a CMOS chip, and electronics on the same chip interface with the integrated electrodes. Impedance tomography maps are reconstructed from measurements with glass microspheres and compared to micrographs, and sources of variation in reconstruction are discussed. This demonstration illustrates the potential of CMOS as a platform where biology, electronics, and potentially photonics can all interact within the same chip.

\section*{Acknowledgment}
The authors thank members of the Caltech High-Speed Integrated Circuits group for useful discussions.



\vspace{12pt}



\end{document}